\documentclass[12pt,preprint]{aastex}
\usepackage{natbib}
\usepackage{graphicx}
\usepackage{amsmath}

\newcommand{\gray}{$\gamma$-ray}
\newcommand{\Kep}{$K_{\rm ep}$}

\newcommand{\caseA}{{Model~Ia}}
\newcommand{\caseAs}{{Model~Ib}}
\newcommand{\caseB}{{Model~IIa}}
\newcommand{\caseBs}{{Model~IIb}}

\newcommand{\hess}{HESS}

\slugcomment{draft version}

\shorttitle{The two-zone origin of {\gray}s from SNR RX~J1713.7$-$3946 }
\shortauthors{Zhang \& Chen}

\begin{document}


\title{The role of the diffusive protons in the $\gamma$-ray emission of supernova remnant RX~J1713.7$-$3946 --- a two-zone model}


\author{ 
  Xiao Zhang\altaffilmark{1}
  and
  Yang Chen\altaffilmark{1,2,3}
}

\altaffiltext{1}{\footnotesize Department of Astronomy, Nanjing University,
163 Xianlin Avenue, Nanjing 210023, China}

\altaffiltext{2}{\footnotesize 
Key Laboratory of Modern Astronomy and Astrophysics,
Nanjing University, Ministry of Education, Nanjing 210023, China}

\altaffiltext{3}{\footnotesize Corresponding author}

\begin{abstract}
RX~J1713.7$-$3946 is a prototype in the \gray -bright supernova remnants (SNRs) and is in continuing debates on its hadronic versus leptonic origin of the \gray\ emission.
We explore the role played by the diffusive relativistic protons that escape from the SNR shock wave in the \gray\ emission, apart from the high-energy particles' emission from the inside of the SNR.
In the scenario that the SNR shock propagates in a clumpy molecular cavity, 
we consider that the \gray\ emission from the inside of the SNR may arise either from the inverse Compton scattering or from the interaction between the trapped energetic protons and the shocked clumps.
The dominant origin between them depends on the electron-to-proton number ratio.
The diffusive protons that escaped from the shock wave during the expansion history can provide an outer hadronic \gray\ component by bombarding the surrounding dense matter.  
The broadband spectrum can be well explained by this two-zone model,
in which the \gray\ emission from the inside governs the TeV band, while the outer emission component substantially contributes to the GeV \gray{s}.
The two-zone model can also explain the TeV \gray\ radial brightness profile that significantly stretches beyond the nonthermal X-ray-emitting region.
In the calculation, we present a simplified algorithm for \citeauthor{Li2010.C}'s (2010) ``accumulative diffusion" model for escaping protons and apply the Markov Chain Monte Carlo method to constrain the physical parameters. 
\end{abstract}





\section{INTRODUCTION}\label{sec:intr}

It is usually believed that Galactic supernova remnants (SNRs) are the main accelerators of cosmic rays (CRs, mainly protons) and can boost particles up to the ``knee" energy of $\sim$3 $\times\ 10^{15}$ eV through the diffusive shock acceleration (DSA).
An important probe of this conjecture is the hadronic \gray\ emission (namely, $\pi^{0}$-decay {\gray}s produced in the inelastic collision between the accelerated protons and the target baryons) from the SNRs that interact with molecular clouds (MCs).
Fortunately, great progress has been made in understanding the SNR paradigm by the detection of a  characteristic $\pi^0$-decay ``bump" in the interacting SNRs IC~443 \citep{Ackermann2013} and W44  \citep{W44.AGILE.2011,Ackermann2013,W44.AGILE.2014}.
Yet, multifaceted investigations of hadronic interaction are still needed to increase and strengthen the evidence for the paradigm in a large sample of SNR-MC association systems \citep[e.g.,][]{Jiang2010.3C397}.
It is, however, often uneasy to distinguish the nature of the \gray{s} between the hadronic scenario and leptonic (inverse Compton [IC] and/or bremsstrahlung) scenario
even in the SNR-MC systems, such as SNR RX J1713.7$-$3946, and deep explorations for the spectral properties and emitting mechanism of them in the GeV--TeV range are essential.

SNR RX~J1713.7$-$3946 (G347.3$-$0.5) was discovered by {\sl ROSAT} X-ray observations \citep[][pp. 267-68]{Pfeffermann1996} and suggested to be the remnant of the historical supernova AD~393 (\citealt{Wang1997.QC}; hereafter an age of the remnant $t_{\rm age}\sim 1620$ yr will be used).
Its shell-like X-ray emission is dominated by a nonthermal component and a lack of thermal line features \citep[e.g.,][]{Koyama1997,Slane1999,Cassam-Chenai2004}.
It is found to be confined in a molecular cavity, and the elevated gas temperature and broad molecular line wings in a few molecular cloudlets are ascribed to the high-energy events of the SNR \citep{Fukui2003,Moriguchi2005}.
The molecular cavity can be reasonably understood as the product of the energetic stellar wind and ionizing photons of the massive progenitor star (which is suggested to be no later than B0--B1 type, with a mass $\ga$15 $\pm\ 2\ M_{\odot}$; \citealt{Chen2013.ZC}).
In the radio band, it shows faint emission and has an average angular diameter of $60^{\prime}$, corresponding to an average radius $R_{s}\, \sim\, 9$ pc at distance $d\, \approx\, 1$ kpc \citep[e.g.][]{Fukui2003,Moriguchi2005}.
It is suggested that the SNR is in the free expansion evolutionary phase \citep[e.g.][]{Moriguchi2005,Sano2010}
and that the forward shock has not yet reached the wall enclosing the wind-blown cavity in view of the high velocity currently measured \citep{J1713.Fermi.2015}.

Since the detection of TeV \gray\ emission from it \citep{Enomoto2002,J1713.HESS.2004,J1713.HESS.2007}, SNR RX~J1713.7$-$3946 has garnered an enormous amount of attention and aroused constant debates on the hadronic versus leptonic origin of its \gray\ emission \citep[e.g.][]{J1713.HESS.2006,Berezhko2008,Liu2008.FFW,Morlino2009a,Zirakashvili2010}.
The debates seemed to have been concluded when the 2\,yr {\sl Fermi}-LAT observations revealed a hard GeV spectrum with a power-law photon index $\Gamma=1.5\pm0.1$ \citep{J1713.Fermi.2011}, which appears to support the leptonic scenario \citep[e.g.,][]{J1713.Fermi.2011, Li2011.LC, Yuan2011.LFB, Ellison2012,Finke2012, Lee2012}.
Subsequently, however, a strong correlation of azimuthal distribution was found between the TeV \gray\ flux and the column density of total interstellar protons, which is in favor of a hadronic origin of the {\gray}s \citep{Fukui2012}. Some hadronic interaction models have also been elaborated to interpret the hard GeV emission \citep{Inoue2012,Gabici2014,J1713.Fermi.2015}. 

Both scenarios in terms of relativistic leptons and protons accelerated via standard DSA can explain the hard spectrum with their advantages, but meanwhile also present difficulties.
In the context of pure leptonic processes, the IC scattering seems to naturally explain the hard GeV \gray\ spectrum without any extra assumptions and can be well compatible with the lack of thermal X-ray line emission.
A leptonic origin of the TeV emission is strongly suggested by \citet{Li2011.LC} using a spectral inversion method independent of the particle acceleration model.
The morphology in TeV band closely matches the nonthermal shell in the radio and X-ray bands \citep{J1713.HESS.2006}, also suggesting an origin from the same electron population.
On the other hand, it is pointed out that single-zone distribution of elections encounters a difficulty in providing a good fit to its broadband spectral energy distribution \citep[SED;][]{Finke2012}.
Also, if the \gray\ emission is leptonic dominated, it remains perplexity why it does not contain any significant hadronic contribution, given the observational fact that the SNR is environed by molecular gas.

In the context of pure hadronic processes, an inhomogeneous shocked medium with dense clumps \citep{Inoue2012,Gabici2014} and a dense shell outside the SNR without contact \citep{J1713.Fermi.2015} are invoked.
In the former case, \citet{Inoue2012} consider that, in the downstream of the shock wave, the penetration depth regulated by energy dependence of Bohm diffusion is generally shorter than the thickness of clumps, which gives the mass of the proton-illuminated gas depending on the energy of incident protons $M(E_{\rm p})\propto E_{\rm p}^{1/2}$ and hence a \gray\ index 0.5 smaller than that of the parent protons.
Alternatively, \citet{Gabici2014} suggest that, because of turbulent layers surrounding the clumps, only protons above a minimum energy can effectively diffuse into clumps, which may result in a hard proton spectrum below that energy in clumps.
It is, however, difficult in this case to explain why no thermal X-ray line emission from the shocked dense medium is observed.
Possibly, the dense clumps swept up by the shock wave can survive and still remain a temperature too low to produce thermal X-rays \citep{Inoue2012,Gabici2014}.
In the latter case, the dense shell (within a thickness of $\sim$0.3 pc) outside the SNR is effectively illuminated by TeV protons in the CR precursor region, giving rise to a hard \gray\ spectrum \citep{J1713.Fermi.2015}.
In this case, the shock does not interact with dense material, and thus the difficulty with the lack of thermal X-rays can be avoided (although the model fluxes in $\sim$10\,GeV--1\,TeV appear to be underestimated).

Here we explore the role played by the diffusive relativistic protons that escape from the SNR shock wave and hit the surrounding dense gas in the \gray\ emission of RX~J1713.7$-$3946, 
considering that the blast wave is propagating in a molecular cavity, in which there may be a mount of shocked clumps. 
Most recently, the TeV \gray\ emission has been found to stretch noticeably beyond the nonthermal X-ray emission region, also indicating a contribution of the escaping protons to the \gray\ emission \citep[][and references therein]{deNaurois2015}.
Currently, there are two types of mechanisms in essence suggested for hadronic emission: $\pi^0$ decay not only can take place in shock-crushed dense clouds \citep[e.g.,][]{Blandford1982,Uchiyama2010,Tang2014.C} but also can occur in the adjacent MCs illuminated by diffusive escaping protons \citep[e.g.,][]{Aharonian1996,Gabici2009,Li2010.C,Ohira2011}.
In the following, we show that the GeV--TeV \gray\ spectrum of RX~J1713.7$-$3946 with updated 5\,yr {\sl Fermi}-LAT data can be perfectly interpreted with a two-zone scenario, in which the $\pi^{0}$-decay \gray{s} resulting from the diffusive protons will substantially contribute to the GeV \gray{s} while 
the leptonic process or the $p$--$p$ interaction in the shocked clumps will be responsible for the TeV \gray{s}.
In the calculation, we present a simplified algorithm for the ``accumulative diffusion" model for escaping protons developed by \citet{Li2010.C} and apply the Markov Chain Monte Carlo (MCMC) method to constrain the model parameters. 

\section{A TWO-ZONE MODEL}

Here we consider two \gray -emitting zones for SNR~RX~J1713.6$-$3946.
The first zone is the cavity wall (the dense matter at the cavity boundary), which is bombarded by the protons that escaped from the shock wave during the history of expansion.
The second zone is inside the SNR, and the emission could be either IC scattering off the
accelerated electrons (\caseA\/) or the hadronic emission from the shocked clumps (\caseB\/).

It is suggested that CR protons that escape from SNRs can illuminate the nearby massive clouds and produce \gray{s} \citep{Aharonian1996}.
Some possible evidence for this scenario has been obtained by the observation of diffuse \gray\ emission from the vicinity of SNR shells, such as in W44 \citep{Uchiyama2012}, with which massive MCs are spatially coincident.
The model has successively been improved and applied to a number of SNR-MC systems to explain the origin of the \gray\ emission \citep[e.g.,][]{Gabici2009,Li2010.C,Li2012.C,Ohira2011,Nava2013}.
In the case of RX~J1713.7$-$3946, although the blast wave may not have hit upon the cavity wall \citep[e.g.][]{Inoue2012}, a considerable fraction of the escaping protons have arrived at the wall.
It can be expected that the $p$--$p$ collision between these protons with the molecules in the wall gives birth to $\pi^0$ decay \gray{s}.
This emission, together with the leptonic emission from the electrons and/or the hadronic emission from the shocked clumps, will be calculated below to jointly interpret the spectral and spatial behavior of the \gray{s} from the SNR.

\subsection{Simplified Algorithm for the Accumulative Diffusion Model}
\label{sec:mod}
For the continuous injection of the accelerated protons that escape from the propagating shock wave surface, \citet{Li2010.C} established an ``accumulative diffusion" model, in which the protons at a given position are a collection of the diffusive protons escaping from the entire shock surface at different radii throughout the history of the SNR expansion. They derived the proton distribution by integrating the solution of the diffusion equation given by \citet{Aharonian1996} for the impulsive point-like source injection \citep[see Equations.\ (1)--(2) in][]{Li2010.C}.
Actually, the algorithm of the model can be simplified by directly solving the diffusion equation of the escaping protons as follows.

Ignoring protons' energy losses, the diffusion equation in the spherically symmetric case is written as
\begin{eqnarray}
\label{eq:dif}
\frac{\partial}{\partial t}f(E_{\rm p},R,t)=\frac{D(E_{\rm p})}{R^{2}}\frac{\partial}{\partial R}R^{2}\frac{\partial}{\partial R}f(E_{\rm p},R,t)+Q(E_{\rm p},R,t),
\end{eqnarray}
where $f(E_{\rm p},R,t)$ is the distribution of the escaping protons with energy $E_{\rm p}$ at a given distance $R$ from the SNR center at a given time $t$ from the explosion time, $Q(E_{\rm p},R,t)$ is the source injection function,  and $D(E_{\rm p})$ is the diffusion coefficient. Here we assume that the diffusion coefficient has the form of $D(E_{\rm p})=10^{28}\chi(E_{\rm p}/10\,{\rm GeV})^{\delta}\ {\rm cm^2}\,{\rm s^{-1}}$, where $\chi$ is the correction factor of slow diffusion around the SNR and $\delta\approx 0.3$--0.7 \citep{Berezinskii1990} is the index of diffusion coefficient.
Following the treatment in \citet{Atoyan1995}, we define a new function $F(E_{\rm p},R,t)=Rf(E_{\rm p},R,t)$, and then Equation\,\eqref{eq:dif} reads as
\begin{eqnarray}
\label{eq:dif2}
\frac{\partial}{\partial t}F(E_{\rm p},R,t)=D(E_{\rm p})\frac{\partial^2}{\partial R^2}F(E_{\rm p},R,t)+Q(E_{\rm p},R,t)\,R.
\end{eqnarray}
The Green function for Eq.\,(\ref{eq:dif2}), namely, the solution for point-like impulsive injection source $Q(E_{\rm p},R,t)R=\delta (R-\xi)\delta (t-t_i)$, is \citep{Atoyan1995}
\begin{eqnarray}
\label{eq:gre}
G(E_{\rm p},R,t;\xi ,t_i) 
    = \frac{1}{\sqrt{\pi} R_{\rm d}}\left\{
      \exp \left[ -\left(\frac{R-\xi }{R_{\rm d}}\right)^2 \right]-
      \exp \left[ -\left(\frac{R+\xi }{R_{\rm d}}\right)^2 \right]
      \right\},
\end{eqnarray}
where $R_{\rm d}=2\sqrt{D(t-t_i)}$ is the diffusion radius. 

The injection function of the protons that escape from the shock front is 
\begin{eqnarray}
\label{eq:inj}
 Q(E_{\rm p},R,t)=\frac{Q_{\rm p}(E_{\rm p})}{4\pi R^{2}}\delta(R-R_s(t)),
\end{eqnarray}
where $Q_{\rm p}(E_{\rm p})$ is the injection rate from the spherical surface $4\pi R^2$ and $R_s(t)$ is the shock radius.
The solution of Equation\,\eqref{eq:dif} is
\begin{equation}\label{eq:sol}
\begin{split}
f(E_{\rm p},R,t) 
         =& \frac{1}{R}\int^{t}_{0}\int^{\infty}_{0}
            Q(E_{\rm p},\xi ,t_i)\xi G(E_{\rm p},R,t;\xi ,t_i)
            {\rm d}\xi {\rm d}t_i  \\
         =& \int^{t}_{0} \frac{Q_{\rm p}(E_{\rm p})}{4\pi^{3/2}R_{s}(t_i) R_{\rm d} R}\left\{
            \exp\left[-\left(\frac{R-R_{s}(t_i)}{R_{\rm d}}\right)^{2}\right]-
            \exp\left[-\left(\frac{R+R_{s}(t_i)}{R_{\rm d}}\right)^{2}\right]
            \right\}{\rm d} t_i.
\end{split}
\end{equation}
This solution, with a single integral, is equivalent to Equation (2) in \citet{Li2010.C}, which is a triple integral, and will be reduced to Equation (8) in \citet{Aharonian1996} for the case of point-like continuous injection as $R_{s}\rightarrow 0$.

\subsection{Model Fit to the Broadband SED}
\label{sec:fit}

The shock-accelerated particles are assumed to obey a power-law distribution with a high-energy cutoff:
\begin{equation}
\label{eq:dis}
dN_{i}/dE_{i}=A_{i}E_{i}^{-\alpha_{i}}\times \exp(-E_{i}/E_{{\rm c},i})
\end{equation}
where $i={\rm e,p}$, $E_{i}$ is the particle kinetic energy, $\alpha_{i}$ is the power-law spectral index and is taken as $\alpha_{\rm e}=\alpha_{\rm p}=\alpha$, and $E_{{\rm c},i}$ is the cutoff energy and will be set to the CR ``knee" energy for the protons, leaving $E_{\rm c,e}$ as a free parameter.
The normalization parameter $A_{i}$ is determined from the total kinetic energy of particles above 1 GeV.

In our calculation, two physical parameters are adopted instead of the normalization parameters: the energy conversion efficiency $\eta$, namely, the faction of explosion energy converted to CR energy, and the number ratio between the accelerated electrons and protons at 1 GeV, $K_{\rm ep}=A_{\rm e}/A_{\rm p}$. 
The energy conversion efficiency ($\eta$) is always difficult to determine directly from the acceleration theory, and there is usually not a consensus in the value of the energy conversion efficiency.
In the {\sl cr}-{\sl hydro}-{\sl NEI} modeling for SNR RX J1713.7$-$3946, it was suggested that about 16\% of the explosion energy has converted into the accelerated particles \citep{Lee2012}. 
A lower value of 6\% was obtained via modeling the \gray\ emission \citep{Gabici2014}.
For electron-to-proton number ratio ({\Kep}), an order of 0.01 was derived from the CRs measured at Earth and a lower limit, $\ga${}$10^{-3}$, was implied according to the radio observations of the SNRs in nearby galaxies \citep{Katz2008}.
 
The distribution of the escaping protons, $f(E_{\rm p},R_{w},t_{\rm age})$, is obtained from Equation\,\eqref{eq:sol}, 
where $R_{w}$ is the radius of the cavity wall and is set to be $\sim$10 pc.
In the calculation, following \citet{Cassam-Chenai2004}, we assume that the SNR radius conforms to the self-similar solution for the free expansion phase \citep{Chevalier1982},
$R_{s}(t)\propto t^{(n-3)/(n-s)}$, where $n$ and $s$ are the power-law indices of the initial outer density profile in the ejecta and of the initial ambient medium density profile. Here we take $n=9$ and $s=0$ (for a shocked wind-dominated medium), but taking $n=12$ or $s=2$ will produce a similar result in our calculation.
We adopt the average remnant radius $R_{s}\approx9$\,pc at present time $t_{\rm age}=1620$\,yr.
The injection rate is given by $5t_{\rm age}Q_{\rm p}(E_{\rm p})=dN_{\rm p}/dE_{\rm p}$.
Based on the calculations of \citep{Lee2012}, we assume that about 20\% of the accelerated protons escape from the SNR. The explosion energy of the SNR is assumed to be $10^{51}$\,erg.

For the hadronic \gray\ emission, we use the analytic photon emissivity developed by \citet{Kelner2006} for $p$--$p$ interaction, including the enhancement factor of 1.84 due to the contribution from heavy nuclei \citep{Mori2009}.
We calculate the leptonic emission following \citet{Blumenthal1970}.
In addition to the cosmic microwave background, the seed photons also include the Galactic interstellar infrared photons at 30 K with an energy density of 0.3 eV\,cm$^{-3}$ \citep[][]{Porter2006}.

Two cases will be considered for the \gray\ emission arising from the inside of the SNR: (I) it is dominated by the IC process of the accelerated electron; and (II) it is dominated by the collision of the accelerated protons with the shocked clumps.

\subsubsection{Case I: Leptonic Emission Dominated Inside}
On the inner side of the shock front, the \gray\ emission is assumed to be mainly from the shock-accelerated electrons, which are also producing the synchrotron emission in the spectral range from radio to X-rays.
The leptonic process in this SNR has often been considered as the origin of the \gray\ emission from it, chiefly owing to the consistency with the hard GeV \gray\ spectrum, the lack of thermal X-ray emission, and its location in a low-density cavity of molecular gas.
In this case, we adopt a typical electron-to-proton number ratio \Kep\ = 0.01.
(With this parameter, the \gray\ emission from the possibly shocked clumps will be insignificant, as can be seen in the following results.)

There are seven parameters in this case (for the combined contribution from the outside diffusive protons and inside electrons; \caseA): $\eta$, $\alpha$, $\delta$, $\chi$, $E_{\rm c,e}$, $B_{\rm SNR}$, and $M_{\rm t}$, where $B_{\rm SNR}$ is the average magnetic field at the shock and $M_{\rm t}$ is the mass of the baryon targets in the cavity wall bombarded by the escaping protons. 
To constrain the model parameters, we employ the MCMC approach, which, based on the Bayesian statistics, is better than the grid approach, with a more efficient sampling of the parameter space of interest, especially for high dimensions.
A brief introduction to the basic procedure of the MCMC sampling can be found in \citet{Fan2010.LYF}, and more details can be found in \citet{Neal1993}, \citet{Gamerman1997}, and \citet{Mackay2003}.
In our MCMC fitting calculation, all parameters are set to be free, but $\delta$ is confined in the range 0.3--0.7, and the observational data with only upper limits are excluded.

\begin{figure}
\centering
\includegraphics[height=54mm,angle=0]{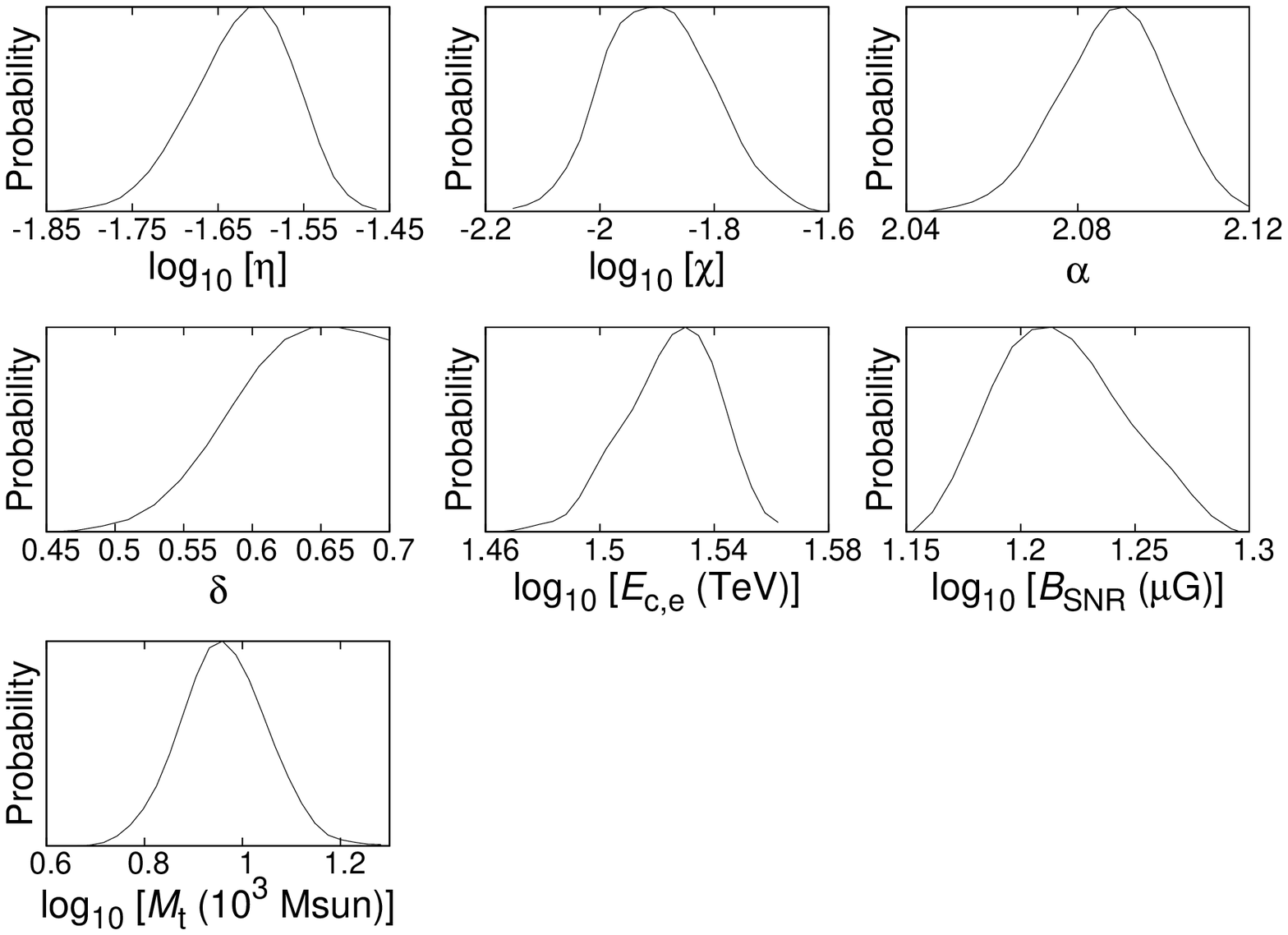}
\includegraphics[height=56mm,angle=0]{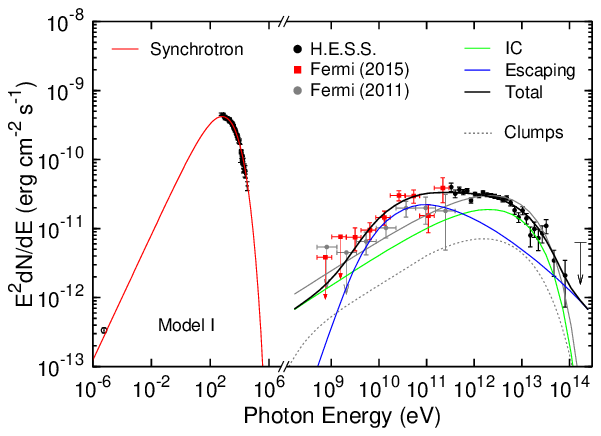}
\caption
{
Fitted results for \caseA.
Left: 1D probability distribution of the parameters; Right: best-fitted broadband SED of SNR RX~J1713.7$-$3946 according to the radio \citep{Acero2009}, X-ray \citep{Tanaka2008}, GeV ({\sl Fermi}: \citealt{J1713.Fermi.2011}, \citealt{J1713.Fermi.2015}), and TeV (\hess: \citealt{J1713.HESS.2011}) data. The gray solid line represents the fitting SED for \caseAs.
}
\label{fig:caseA}
\end{figure}

The one-dimensional (1D) probability distributions of model parameters and the best-fit SED are shown in Figure~\ref{fig:caseA}.
The 1D probability distribution of each parameter converges well with a single peak (left panel of Figure~\ref{fig:caseA}).
The best-fit parameters with 1$\sigma$ statistical uncertainties and the $\chi^2$ values are listed in Tables~\ref{tab:params} and Table~\ref{tab:chi2}, respectively.
The model \gray\ SED (see right panel of Figure~\ref{fig:caseA}) shows that this model perfectly explains the \gray\ data.
In the energy range $\sim$3--500 GeV, the \gray{s} are mainly contributed by the hadronic emission with a peak flux therein; in the remaining regime, they are dominated by the leptonic emission.
The fluxes above 0.25\,TeV of the hadronic (outer) and leptonic (inner) components are about $5.3\times10^{-11}\, {\rm erg\,cm^{-2}\,s^{-1}}$ and $7.6\times10^{-11}\, {\rm erg\,cm^{-2}\,s^{-1}}$, respectively.
The fluxes in the range from 0.3 to 40\,TeV of the outer and inner components are about $4.7\times10^{-11}\, {\rm erg\,cm^{-2}\,s^{-1}}$ and $7.3\times10^{-11}\, {\rm erg\,cm^{-2}\,s^{-1}}$, respectively.
For the escape process, we got $\chi\sim0.01$, which is similar to the previous estimates for SNR environs \citep[e.g.,][]{Fujita2010,Li2012.C,Ohira2012,Tang2015.C}, and $\delta\sim0.65$, which is in the normal range.
The mass of the baryon targets, $8.8\times10^3 M_{\odot}$, seems reasonable as compared to the estimate of the surrounding interstellar protons reservoir, $2\times10^{4} M_{\odot}$ \citep{Fukui2012}.
In addition, the \gray\ emission from the possibly shocked clumps (see the next subsection for description) in this case is also plotted by gray dotted line in Figure~\ref{fig:caseA}, implying an insignificant contribution to the total \gray\ emission.

For comparison, we also test the notion of the pure leptonic process as the origin of the GeV--TeV \gray{s}, incorporating the updated {\sl Fermi} data (\caseAs).
The best-fit SEDs are displayed by the gray solid line in Figure~\ref{fig:caseA}.
The model parameters and $\chi^2$ values of the best fit are also listed in Tables~\ref{tab:params} and \ref{tab:chi2}, respectively. The $\chi^2$ values in various wavebands are obviously higher than those in \caseA\ except in the radio. The best-fit SED shows that the pure leptonic model underestimates the GeV flux, similar to the results of previous multiband fits with simple leptonic models \citep[e.g.,][]{Fan2010.LYF,Finke2012}.
In addition, we also calculate the $\chi^2$ value in the GeV band using our best-fit SED and the previous {\sl Fermi} data \citep{J1713.Fermi.2011}, which are listed in Table~\ref{tab:chi2} in parentheses.

\subsubsection{Case II: Hadronic Emission Dominated Inside}

In this case, the \gray\ emission in the downstream are mainly from the decay of the neutral pions generated in the inelastic collision between the relativistic protons and shocked dense clumps.
It is pointed out that there may be some dense clumps shocked by the blast wave, which can act as a target for the bombardment of energetic protons \citep[e.g.][]{Moriguchi2005,Sano2010,Sano2013}.
As mentioned in section Section \ref{sec:intr}, a few hadronic models have been proposed to interpret the observed GeV--TeV \gray\ spectrum.
The IC \gray\ generation may be suppressed owing to the high magnetic field amplified by the turbulence induced by the shock--cloud interaction \citep[e.g.][]{Inoue2012}.
This case requires a relatively small amount of electrons (for the synchrotron), and thus we adopt a low value of the electron-to-proton number ratio, $K_{\rm ep}=10^{-3}$.

Following \citet{Inoue2012}, the total mass of the gas in the clumps directly bombarded by the accelerated protons, $m_{\rm c,tot}$, can be obtained via three steps: 
(1) Estimate the average mass density $\rho_{\rm c}$ for each clump with physical parameters given by \citet[][see their Tables~3 and 4]{Sano2013}.
(2) Estimate the mass in the outer layer of each clump that is penetrated by relativistic protons using $m_{\rm c}=4\pi \rho_{\rm c} a_{\rm c}^2\,l_{\rm pd}$ \citep[see][]{Inoue2012}, where $a_{\rm c}$ is the clump radius and $l_{\rm pd}$ is the penetration depth of the accelerated protons into the clumps during diffusion time $t_{\rm pd}$.
The diffusion time is estimated as $t_{\rm pd}\sim 0.2\, (l_{m}/1\,{\rm pc})(V_s/4000\,{\rm km\,s^{-1}})^{-1}\,{\rm kyr}$, where $V_s=4000\,{\rm km\,s^{-1}}$ \citep[e.g.,][]{Cassam-Chenai2004,Acero2009} is the shock velocity and $l_{m}\sim$\,1\,pc is the mean length that the shock moves forward from the clumps, estimated from the spatial distribution of the clumps (Sano et al.\ 2013).
(3) Scale the mass of the gas collided by protons according to the area fraction covered by X-rays for each clump and sum the mass of the clumps.
Hence, we obtain 
\begin{equation}
m_{\rm c,tot}\approx 13\,\left(\frac{\eta_{\rm B}}{1}\right)
               \left( \frac{E_{\rm p}}{10\,{\rm GeV}} \right)^{1/2}
               \left( \frac{B_{\rm MC}}{100\,\mu {\rm G}} \right)^{-1/2}
               \left( \frac{t_{\rm pd}}{0.2\,{\rm kyr}} \right)^{1/2} M_{\odot},
\end{equation}
where $\eta_{\rm B}$ is the degree of magnetic field fluctuations and $B_{\rm MC}$ is  the magnetic field in the dense region of MCs \citep[see][]{Inoue2012}.
We assume that the trapped accelerated protons which uniformly distribute in the downstream have a power-law energy distribution, which can also be described by Equation~\eqref{eq:dis}.
In order to account for the \hess\ data, however, the proton cutoff energy needs to be a free parameter to be decided.

\begin{figure}
\centering
\includegraphics[height=54mm,angle=0]{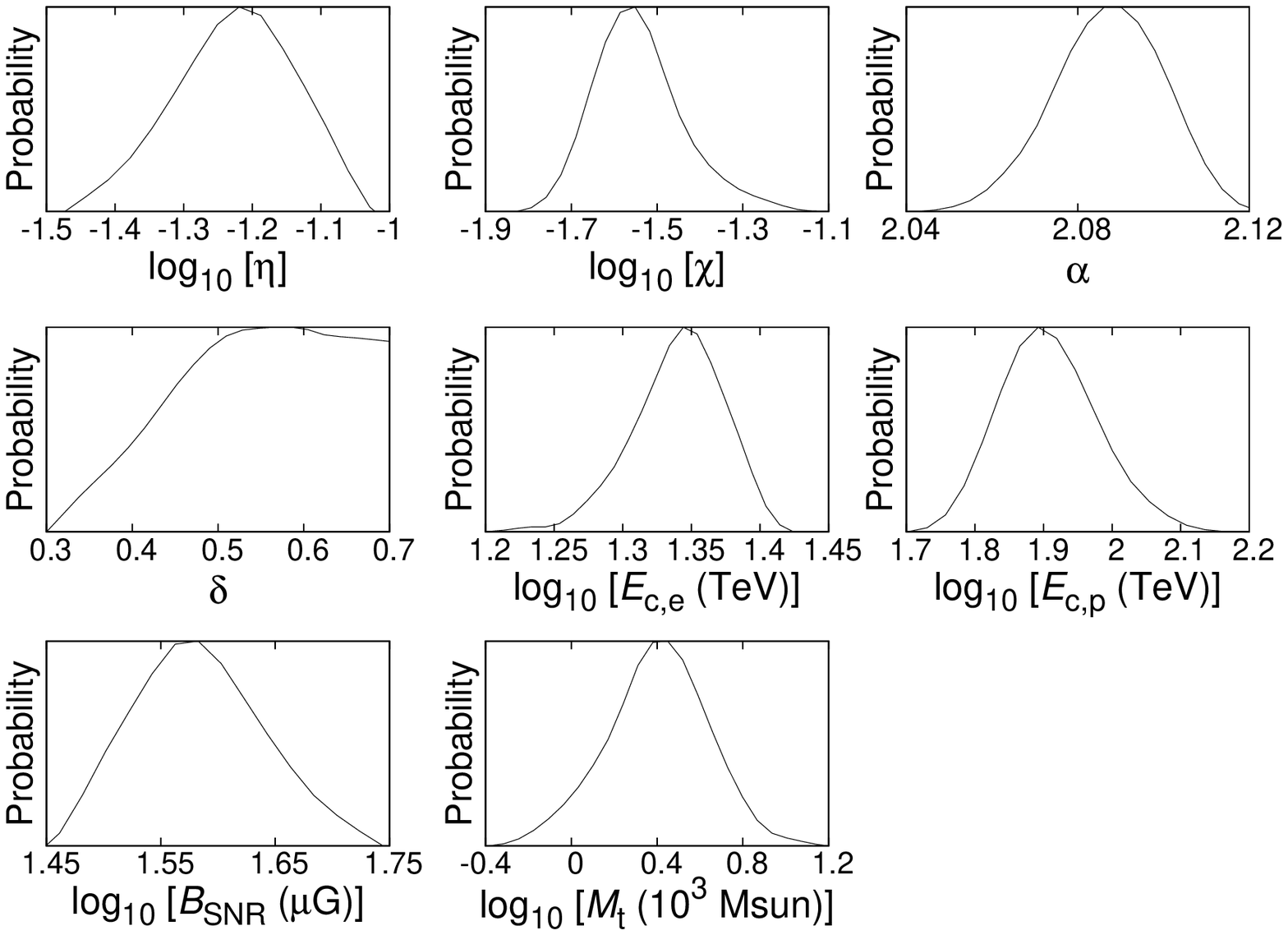}
\includegraphics[height=56mm,angle=0]{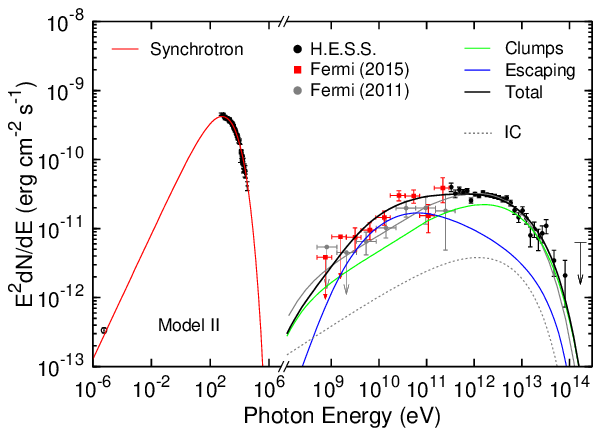}
\caption
{
Same as in Figure~\ref{fig:caseA} but for \caseB. The gray solid line in the right panel represents the fitting SED for \caseBs.
}
\label{fig:caseB}
\end{figure}

There are eight parameters for the combined contribution from the outside diffusive protons and inside trapped protons (\caseB): $\eta$, $\chi$, $\alpha$, $\delta$, $B_{\rm SNR}$, $E_{\rm c,e}$, $E_{\rm c,p}$, and $M_{\rm t}$, which are constrained by using the MCMC method.
The 1D probability distributions of model parameters and the best-fit SED are shown in Figure~\ref{fig:caseB}.
The best-fit parameters with 1$\sigma$ statistical uncertainties and the $\chi^2$ values are also listed in Tables~\ref{tab:params} and \ref{tab:chi2}, respectively.
Similar to that in \caseA, the model \gray\ SED (see right panel in Figure~\ref{fig:caseB}) also is the sum of two components, the \gray\ emission from the shocked clumps penetrated by the trapped protons and the cavity wall illuminated by the escaping protons.
In the energy range $\sim$2--200 GeV, the \gray{s} are mainly contributed by the escaping protons with a peak flux therein; in the remaining range, they are dominated by the contribution of the trapped protons.
The fluxes above 0.25\,TeV from the escaping (outer) and trapped (inner) protons are about $3.4\times10^{-11}\, {\rm erg\,cm^{-2}\,s^{-1}}$ and $8.7\times10^{-11}\, {\rm erg\,cm^{-2}\,s^{-1}}$, respectively.
The 0.3--40\,TeV fluxes for the outer and inner components are about $3.1\times10^{-11}\, {\rm erg\,cm^{-2}\,s^{-1}}$ and $8.4\times10^{-11}\, {\rm erg\,cm^{-2}\,s^{-1}}$, respectively.
Parameters $\chi$ and $\delta$ (for the escaping protons) are similar to those in \caseA.
The IC \gray\ emission is trivial in this case, as displayed with gray dotted line in Figure~\ref{fig:caseB}.

We also consider the downstream proton--clump interaction as the only mechanism of the GeV--TeV \gray\ emission (referred to \caseBs), 
with the spectral fit parameters also listed in Tables~\ref{tab:params} and \ref{tab:chi2}.
The best-fit SED is displayed by the gray solid line in Figure~\ref{fig:caseB} and shows that the contribution from the interaction between the trapped protons and the shocked clumps alone will also underestimate the GeV flux.

\section{DISCUSSION}

As stated above, although the pure IC scattering or the pure shock--clump interaction inside the SNR can explain the low spectral index at GeV band, such simple models tend to underestimate the GeV-band flux.
In view of the expansion of the SNR in a molecular cavity and the presence of the dense molecular gas surrounding the SNR, we additionally take into account the role of the energetic protons that escaped from the blast shock during the expansion history.
These protons can contribute the \gray\ emission due to ``illumination" of the nearby dense gas due to $p$--$p$ collision.
Considering that the SNR shock is still propagating in the low-density medium (in which there may be interspersal dense clumps), we have calculated this illumination effect on the broadband spectrum of SNR RX~J1713.7$-$3946 in a two-zone model.

\subsection{Spectral Behavior of the Gamma-ray Emission}

From the fitted SED and the $\chi^2$ value, we can see that both \caseA\ and \caseB\ give similarly good results and the hadronic \gray\ emission from the escaping protons plays an important role in the explanation of the spectral data, especially in the GeV band.
The flux of this component, peaking around $\sim80$\,GeV, naturally compensates the underestimate at the GeV band in the aforementioned simple models.
From the point of view of spectral shape, the hadronic spectrum from the inside clumps (in \caseB) actually has a similar effect on the total \gray\ emission to the leptonic one (in \caseA).
Given the presence of the shocked clumps, the dominant mechanism of \gray\ emission in the downstream depends on the electron-to-proton number ratio, \Kep: it was found that the hadronic emission from the inside clumps dominates for \Kep\,$<\,\sim$4$\times10^{-3}$ while the leptonic component dominates for \Kep\,$>\,\sim$4$\times10^{-3}$.
Both models give a small energy transfer fraction $\eta<10\%$ for this young SNR, similar to those suggested for SNRs Cas~A \citep[e.g.][]{Cas-A.Fermi.2010} and Tycho \citep[e.g.][]{Zhang2013.CLZ}.

The two models entail different cutoff energies $E_{\rm c,p}$ of shock-accelerated protons.
It is constrained as $81^{+14}_{-12}$\,TeV by the current \hess\ data in \caseB, while it is fixed as the ``knee" energy in \caseA.
This leads to the difference of the \gray\ spectrum above $100$\,TeV between the two cases (see Figure~\ref{fig:caseA} and \ref{fig:caseB}).
If we consider Bohm diffusion in the acceleration process, the maximum energy of the shock-accelerated protons can be estimated as $E_{\rm m,p}\sim1\times10^3\,(B_{\rm SNR}/38\,{\mu {\rm G}})(V_{s}/4000\,{\rm km s^{-1}})^2(t_{\rm age}/1.6\,{\rm kyr})$\,TeV \citep[e.g.][]{Ohira2012}, which is close to the ``knee" energy and obviously higher than the fitted value in \caseB.
Therefore, the \gray\ observation above 100\,TeV (e.g., the LHAASO project) is expected to give a constraint on this parameter and then possibly make a distinction between the two cases.

\subsection{Spatial Behavior of the Gamma-ray Emission}

Our models imply two emission zones, but, unfortunately, the site of $\sim$GeV emission at the cavity wall cannot be resolved by the {\sl Fermi}-LAT owing to the large point spread function (PSF). 
However, the spatial distribution of the two-zone emission can be partly reflected in the radial surface brightness profile of the TeV emission.
With the help of the angular resolution (better than $3'$) and the increased data set of 150 hr of \hess\ observation, it is found that the TeV \gray\ emission appears significantly more extended than the X-ray one \citep[][and references therein]{deNaurois2015}.
It may provide evidence of escape of protons that would emit hadronic \gray{s} when interacting with surrounding matter \citep{deNaurois2015,Lemoine2015}.

\begin{figure}
\centering
\includegraphics[height=56mm,angle=0]{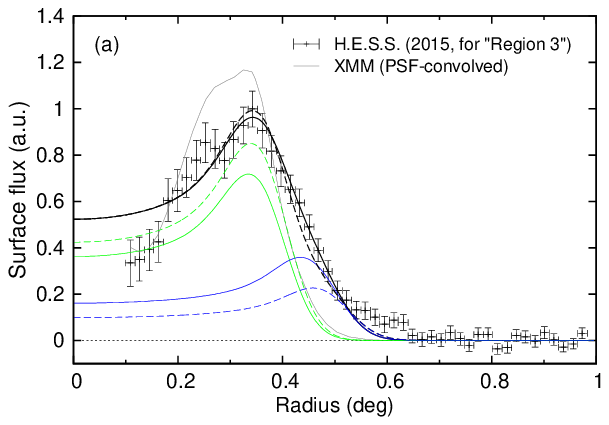}
\includegraphics[height=56mm,angle=0]{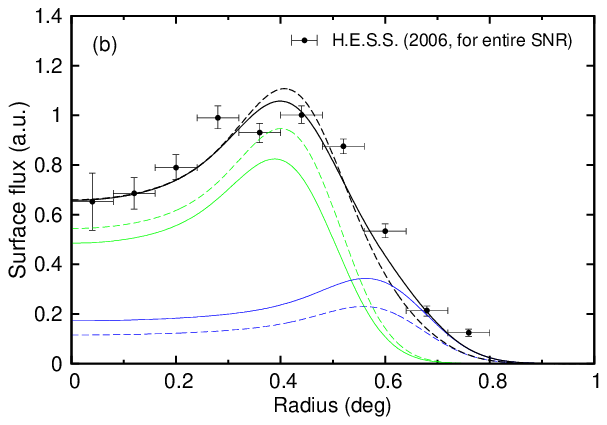}
\caption
{
(a) TeV \gray\ radial surface brightness profile of sector ``region 3." 
The \hess\ flux data and the {\sl XMM-Newton} X-ray brightness profile convolved with the \hess\ PSF of $3'$ (gray line) are adopted from \citet{deNaurois2015}. 
The radial profile of ``region 3" is fitted with a two-zone emission (in black),
which is the sum of the inner (in green) and outer (in blue) components.
The solid and dashed lines represent the profiles in \caseA\ and \caseB, respectively.
All of the fitting profile lines have been convolved with the \hess\ PSF of $3'$.
(b) TeV \gray\ radial surface brightness profile for the entire remnant. The \hess\ flux data are adopted from \citet{J1713.HESS.2006}. The meanings of all lines convolved with the \hess\ PSF of $5'$ are the same as those in (a).
}
\label{fig:pro}
\end{figure}

Here we present a crude comparison of the observed TeV \gray\ radial brightness profile with that generated from our two-zone model.
We fit the TeV profile (above 0.25\,TeV) of the sector-shaped ``region 3" given in \citet{deNaurois2015}, assuming uniform emissivities in the two emission zones.
For \caseA, a flux ratio (above 0.25\,TeV) between the outer and inner components of $\sim$4.1:5.9 is adopted from the above model calculation for the \gray\ emission of the entire remnant.
In Figure~\ref{fig:pro}(a), the fitting profile line (black solid line, which is the sum of the leptonic [green solid line] and hadronic [blue solid line] components) can reproduce both the TeV brightness peak, which is coincident with the nonthermal X-ray peak, and the broad profile wing that extends outside the X-ray-emitting region.
This fit gives the inner radius of the cavity wall $R_{w}\approx 1.13R_{s}$ and the thicknesses of both the outer and inner emission regions $\approx0.12R_{s}$, where $R_{s} \approx7$\,pc (or $24'$) for ``region 3."
The fitted $R_s\approx24'$ of ``region 3" is consistent with the outer edge of the southwestern X-ray arc (in ``region 3") shown in the {\sl XMM-Newton} and {\sl Suzaku} images \citep{Acero2009,Sano2013}.
This profile fit for ``region 3" illustrates that the outstretched \gray\ emission can be ascribed to a contribution from an outer emitting region.

For \caseB, the \gray\ emission in the downstream is suggested to mainly come from the shocked clumps. 
According to Sano et al. (2013), most of the interacting clumps are distributed on the boundary of the SNR.
Although their 3-dimensional spacial distribution may be complicated, for simplicity, we assume that they are all close to the SNR edge.
According to the flux ratio (above 0.25\,TeV) between the outer (escape) and inner (trapped) components, $\sim$2.8:7.2, we fit the TeV profile, as in \caseA, and obtain the inner radius of the cavity wall $R_{w}\approx 1.20R_{s}$ and the thicknesses of the outer and inner emission regions $\approx0.09R_{s}$ and $\approx0.10R_{s}$, respectively.
The fitting profile line is plotted by the black dashed line in Figure~\ref{fig:pro}(a) and also can reproduce the observed features, including the outer extending part.

Similar fittings to the earlier 0.3--40\,TeV \hess\ data of the entire SNR \citep{J1713.HESS.2006} are also made with \caseA\ and \caseB\ (see Figure~\ref{fig:pro}(b)). 
For \caseA, according to the 0.3--40 TeV flux ratio between the outer and inner components, $\sim$3.9:6.1, we obtain 
the thicknesses of the outer and inner emission zones $\approx0.22R_{s}$ and $\approx0.23R_{s}$, respectively.
For \caseB, based on the 0.3--40\,TeV flux ratio between the two zones, $\sim$2.7:7.3,
the thicknesses of the two zones are fitted as $\approx0.20R_{s}$ and $\approx0.18R_{s}$, respectively.
For both models, $R_{s}=9$\,pc and $R_{w}=10$\,pc are used.
The fitting profiles match the observational data as well.

A dense shell located just outside the shock front without contact has recently been suggested to explain the GeV--TeV \gray\ emission via a $p$--$p$ interaction mechanism \citep{J1713.Fermi.2015}.
In their model, the \gray\ emission arises from the CR precursor region, where the accelerated protons directly interact with the target particles without an escape process.
Both GeV and TeV \gray\ emission are dominated by hadronic emission and should be outside the remnant. 
Especially for TeV \gray{s}, the emission is mainly concentrated in the dense shell with a thickness of 0.3\,pc (corresponding to $\sim$1$'$ at a distance of 1\,kpc), which may be difficult to match the broad TeV radial profile even if the PSF of \hess\ and the projected effect are taken into account.
In our model, the emission arises from two zones: the emission from the inside of the SNR can be either due to IC scattering or collision of the trapped protons with the shocked clumps, and the outer emission is ascribed to the collision of the protons that escaped from the shock front during the previous expansion with the surrounding dense matter. The outer emission mainly contributes in the GeV band and the inner emission dominates in the TeV band.

\section{CONCLUSION}
Motivated by the intriguing GeV--TeV \gray\ spectrum of SNR RX~J1713.7$-$3946 and the TeV \gray\ radial profile, 
we have explored the role played by the diffusive relativistic protons that escaped from the SNR shock wave in the \gray\ emission, apart from the high-energy particles' emission from the inside of the SNR.
In the scenario that the SNR shock is propagating in a molecular cavity, in which there may be interspersal shocked dense clumps,
we consider that the \gray\ emission from the inside of the SNR may arise either from the IC scattering (\caseA\/) or from the interaction between the trapped energetic protons and the clumps (\caseB\/).
The dominant origin between them depends on the electron-to-proton number ratio, above or below $\sim$4$\ \times\ 10^3$. 
The surrounding molecular cavity wall is considered to also produce \gray\ emission due to the ``illumination" by the diffusive protons that escaped from the shock wave during the expansion history.
We simplify the algorithm for the \citet{Li2010.C} ``accumulative diffusion" model for diffusive escaping protons. This two-zone model is fit to the broadband spectrum of the SNR that incorporates the updated 5\,yr Fermi data, with application of the MCMC method.
The broadband fluxes can be well explained by the two-zone model, in which the \gray\ emission from the inside governs the TeV band, while the outer emission component substantially contributes to the GeV \gray{s} and naturally compensates for the underestimate of the GeV flux of the inner component.
In the Meantime, we show that the two-zone model can also reproduce the TeV \gray\ radial brightness profile that, at a resolution better than $3'$, significantly extends outside the nonthermal X-ray-emitting region.



\acknowledgments
X.Z.\ is indebted to Hui Li for the helpful discussion on the accumulative diffusion model for escaping protons, to Qiang Yuan for providing the MCMC code, which is adapted from the COSMOMC package \citep{Lewis2002},
and to Ke-ping Qiu and Yang Su for the helpful discussion.
We thank the support of NSFC grant 11233001, 973 Program grant 2015CB857100, grant 20120091110048 from the Educational Ministry of China, and the program B for Outstanding PhD candidate of Nanjing University.

\begin{deluxetable}{llcccccccc}
\tabletypesize{\scriptsize}
\tablecaption{Fitted Parameters with 1$\sigma$ Statistical Error.\label{tab:params}}
\tablewidth{0pt}
\tablehead
{
Model & \colhead{$K_{\rm ep}$} & \colhead{$\eta$} & \colhead{$\alpha$} & \colhead{$B_{\rm SNR}$} & \colhead{$E_{{\rm c,e}}$} & \colhead{$E_{{\rm c,p}}$} & \colhead{$\chi/0.01$}  & \colhead{$\delta$} &  \colhead{$M_{\rm t}$}\\
   &  &(\%) &  & (${\mu}$G) & (TeV) & (TeV) &  &  & ($10^{3}\, M_{\odot}$)
}
\startdata
Ia & $0.01^{a}$  & $ 2.5^{+0.6}_{-0.9}$ & $2.09^{+0.03}_{-0.04}$ & $15.9^{+4.2}_{-1.3}$ &
        $34.2^{+1.7}_{-4.0}$ & $\cdots$ & $1.2^{+1.0}_{-0.4}$  & $0.67^{+0.03}_{-0.14}$ & $ 8.8^{+6.6}_{-3.1}$ \\
Ib & $0.01^{a}$  & $ 4.1^{+0.5}_{-0.5}$ & $2.11^{+0.02}_{-0.02}$ & $12.7^{+0.3}_{-0.3}$ &
        $38.6^{+0.7}_{-0.8}$ & $\cdots$ & $\cdots$ & $\cdots$ & $\cdots$\\
IIa \ & $0.001^{a}$ & $ 6.0^{+2.9}_{-3.3}$ & $2.09^{+0.08}_{-0.06}$ & $37.9^{+6.2}_{-5.5}$ &
        $22.2^{+2.9}_{-5.0}$ & $81.3^{+13.7}_{-11.9}$ & $ 3.1^{+0.7}_{-0.4}$ & $0.55^{+0.15}_{-0.25}$ &
        $2.6^{+1.8}_{-1.0}$  \\
IIb\ & $0.001^{a}$ & $10.6^{+1.8}_{-1.5}$ & $2.10^{+0.02}_{-0.03}$ & $27.6^{+1.5}_{-2.0}$ &
        $26.2^{+1.0}_{-0.8}$ & $60.4^{+08.8}_{-10.1}$ & $\cdots$ & $\cdots$ & \nodata
\enddata
\tablenotetext{a}{Fixed in the model fit.}
\end{deluxetable}

\begin{deluxetable}{llccccc}
\tabletypesize{\scriptsize}
\tablecaption{Best-fit $\chi^2$ value for Each Set of Data.\label{tab:chi2}}
\tablewidth{0pt}
\tablehead
{
Model  & \colhead{Radio} & \colhead{X-Ray} & \colhead{GeV} & \colhead{TeV} & \colhead{$\chi^2_{\nu}({\rm d.o.f.})$}
}
\startdata
Ia    &  3.8 & 364.6 & 10.2 (13.5$^{a}$) &  37.1 & 1.78(233) \\
Ib    &  0.1 & 374.1 & 36.3 ( 5.5$^{a}$) & 120.0 & 2.25(236) \\
IIa   &  3.7 & 364.6 &  9.1 ( 7.0$^{a}$) &  36.4 & 1.78(232) \\
IIb   &  0.8 & 368.6 & 23.9 ( 3.0$^{a}$) &  41.7 & 1.85(235)  
\enddata
\tablenotetext{a}{Corresponding to the old {\sl Fermi} data \citep{J1713.Fermi.2011}.}
\end{deluxetable}

\end{document}